# Incommensurate-commensurate magnetic phase transition in the double tungstate Li$_2$Co(WO$_4$)$_2$*

Xiyu Chen[1], Ning Ding[2], Meifeng Liu[1,†], Tao Zou[3,‡], V. Ovidiu Garlea[4], Jingwen Gong[1], Fei Liu[1], Yunlong Xie[1], Lun Yang[1], Shuhan Zheng[1], Xiuzhang Wang[1], Shuai Dong[2,§], T. Charlton[4], Jun-Ming Liu[1,5]

[1] *Institute for Advanced Materials, Hubei Normal University, Huangshi 435002, China*

[2] *School of Physics, Southeast University, Nanjing 211189, China*

[3] *Collaborative Innovation Center of Light Manipulations and Applications, Shangdong Normal University, Jinan 250358, China*

[4] *Neutron Scattering Division, Oak Ridge National Laboratory, Oak Ridge, Tennessee 37831, USA*

[5] *Laboratory of Solid State Microstructures and Innovative Center of Advanced Microstructures, Nanjing University, Nanjing 210093, China*

**Abstract:**
Magnetic susceptibility, specific heat, and neutron powder diffraction measurements have been performed on polycrystalline Li$_2$Co(WO$_4$)$_2$ samples. Under zero magnetic field, two successive magnetic transitions at $T_{N1}$ ~ 9.4 K and $T_{N2}$ ~ 7.4 K are observed. The magnetic ordering temperatures gradually decrease as the magnetic field increases. Neutron diffraction reveals that Li$_2$Co(WO$_4$)$_2$ enters an incommensurate magnetic state with a temperature dependent ***k*** between $T_{N1}$ and $T_{N2}$. The magnetic propagation vector locks-in to a commensurate value ***k*** = (½, ¼, ¼) below $T_{N2}$. The antiferromagnetic structure is refined at 1.7 K with Co$^{2+}$ magnetic moment 2.8(1) μ$_B$, consistent with our first-principles calculations.

* Project supported by the National Natural Science Foundation of China (Grant Nos. 11834002, 12074111, 11704109), the National Key Research Projects of China (Grant No. 2016YFA0300101).

† Corresponding author. E-mail: lmfeng1107@hbnu.edu.cn

‡ Corresponding author. E-mail: taozoucn@gmail.com

§ Corresponding author. E-mail: sdong@seu.edu.cn

# 1. Introduction

Frustrated magnetic systems, especially those with low-dimensional characteristics, have drawn considerable attention due to their exotic magnetic ground states and novel quantum phenomena originating from strong quantum fluctuations.[1-5] In ideal low-dimensional antiferromagnets, three-dimensional long-range magnetic ordering does not form even at absolute zero temperature.[6] However, the three-dimensional long-range magnetic orders have been observed in most low-dimensional antiferromagnets at finite temperatures due to weak interchain or interlayer couplings under zero magnetic field (or an external critical magnetic field).[7-11] Besides, frustration also plays a crucial role in these magnetic systems. Frustrated magnetic interactions can enhance the spin fluctuations which suppress magnetic ordering temperatures.[12,13] Further, it potentially engenders non-collinear magnetic ground states and induces spontaneous ferroelectric polarization.[14,15] In recent years, cobalt-based frustrated magnets have been extensively explored due to their fascinating physics, such as field-induced order-disorder transition and quantum criticality in Ising-like screw chain $SrCo_2V_2O_8$ and $BaCo_2V_2O_8$,[16-19] 1/3 quantum magnetization plateau in $Ba_3CoSb_2O_9$,[20] spin-driven multiferroicity in $Ba_3CoNb_2O_9$,[15] quantum spin liquid states in Co-based triangular lattice $Na_2BaCo(PO_4)_2$,[21] and potential field-induced Kitaev quantum spin liquid in $BaCo_2(AsO_4)_2$.[22] The discovery of such strange magnetic behaviors and novel magnetic ground states have evoked the research interests in more Co-based magnets.[23-25]

In this work, another Co-based magnet will be studied. $Li_2Co(WO_4)_2$ was reported to possess two successive antiferromagnetic (AFM) transitions at $T_{N1}$ ~ 9 K and $T_{N2}$ ~ 7 K, and the magnetic susceptibility displays a broad maximum associated with short-range spin order around 11 K.[26] This system belongs to the double tungstates family $Li_2M(WO_4)_2$ ($M$ = Co, Ni, Cu) transition metal oxides where magnetic ions form quasi-triangular lattices and, thus, the strong frustrated magnetism could be observed in this family.[26-29] Recently, nuclear magnetic resonance (NMR) and neutron powder diffraction (NPD) measurements were carried out on the sister compound Li$_2$Ni(WO$_4$)$_2$. It has been confirmed that this compound enters the incommensurate SDW-type (spin-density-wave) state below 18 K, followed by a commensurate AFM state with the propagation vector $k$ = (½, 0, ½) below 12.5 K.[27] In contrast, the other known member of this series, Li$_2$Cu(WO$_4$)$_2$, undergoes a single AFM transition below approximately 3.9 K to a collinear AFM state defined by the propagation vector $k$ = (0, ½, 0).[29] Meanwhile, another family of double tungstates $A^I B^{III}$(WO$_4$)$_2$ ($A$ = alkali metal, $B$ = trivalent cation or rare-earth element) has also been extensively studied.[30-34] Similar to Li$_2$Co(WO$_4$)$_2$, LiFe(WO$_4$)$_2$ also undergoes two sequential AFM phase transitions.[32] Interestingly, LiFe(WO$_4$)$_2$ was confirmed as type-II multiferroic material.[32] Recently, we reported a study on double molybdates LiFe(MoO$_4$)$_2$.[35] As well as Li$_2M$(WO$_4$)$_2$ ($M$ = Co, Ni, Cu), LiFe(MoO$_4$)$_2$ belongs to the triclinic space group $P$-1 (No.2). NPD revealed that it orders with a commensurate propagation vector $k$ = (0, ½, 0).[35] The magnetic structure of Li$_2$Co(WO$_4$)$_2$ has remained unclear.

In this work, we will focus on the determination of magnetic structure of Li$_2$Co(WO$_4$)$_2$ by means of neutron powder diffraction and theoretical calculations. Li$_2$Co(WO$_4$)$_2$ undergoes two successive magnetic phase transitions at $T_{N1}$ ~ 9.4 K and $T_{N2}$ ~ 7.4 K. Neutron diffraction reveals that Li$_2$Co(WO$_4$)$_2$ enters an incommensurate magnetic state with varying $k$ between $T_{N1}$ and $T_{N2}$. The magnetic propagation vector locks-in to a commensurate value $k$ = (½, ¼, ¼) below $T_{N2}$. The AFM structure is refined at 1.7 K with Co$^{2+}$ magnetic moment 2.8(1) μ$_B$, which is further confirmed by our first-principles calculations.

## 2. Methods

Polycrystalline Li$_2$Co(WO$_4$)$_2$ samples were synthesized using the conventional solid-state reaction method. Highly purified Li$_2$CO$_3$, CoO, and WO$_3$ powder were mixed and ground in a stoichiometric ratio of 1:1:2 and fired at 550 ℃ for 24 h in air. The resultant powder samples were reground and pressed into pellets and heated at 650 ℃ for 24 h in air. The phase purity of Li$_2$Co(WO$_4$)$_2$ were checked using X-ray diffraction (XRD) with Cu $K$α radiation at room temperature (SmartLab Se, Rigaku).

The magnetic susceptibility $\chi(T)$ under different applied magnetic fields ($H$) were measured using Physical Property Measurement System (PPMS, Quantum Design) in zero-field cooling (ZFC) and field cooling (FC) modes. The specific heat ($C_p$) was measured using the heat relaxation method on PPMS. NPD measurements were carried out at the HB2A diffractometer, ORNL.

The first-principle calculations were performed on the basis of spin-polarized density-functional theory (DFT) implemented in Vienna *ab initio* Simulation Package (VASP) code.[36,37] For the exchange-correlation functional, the Perdew-Burke-Ernzerhof for solids function (PBEsol) of the generalized gradient approximation (GGA) was used[38] and the Hubbard $U$ ($U_{eff}$ = 4 eV[39]) was applied using the Dudarev parametrization.[40] The energy cutoff was fixed at 600 eV, and the W's $5p6s5d$ electrons were treated as valence states. All geometries were optimized until none of the residual Hellmann-Feynman forces exceeded 0.005 eV/Å.

## 3. Results and discussion

$Li_2Co(WO_4)_2$ crystallizes in the triclinic space group $P$-1 (No.2), as shown in Fig. 1(a). Adjacent $CoO_6$ octahedra are indirectly connected by $WO_5$ pyramids. The XRD results confirm the high-quality of our samples, as plotted in Fig. 1(c). The refined lattice parameters of $Li_2Co(WO_4)_2$ are $a$ = 4.9247(03) Å, $b$ = 5.6707(38) Å, $c$ = 5.8858(63) Å, $\alpha$ = 69.481(0)°, $\beta$ = 91.462(2)°, $\gamma$ = 116.141(4)° ($\chi^2$ = 1.71, $R_p$ = 4.86%, $R_{wp}$ = 6.21%), in consistent with previous studies.[26,41] The detailed lattice parameters are presented in Table 1.

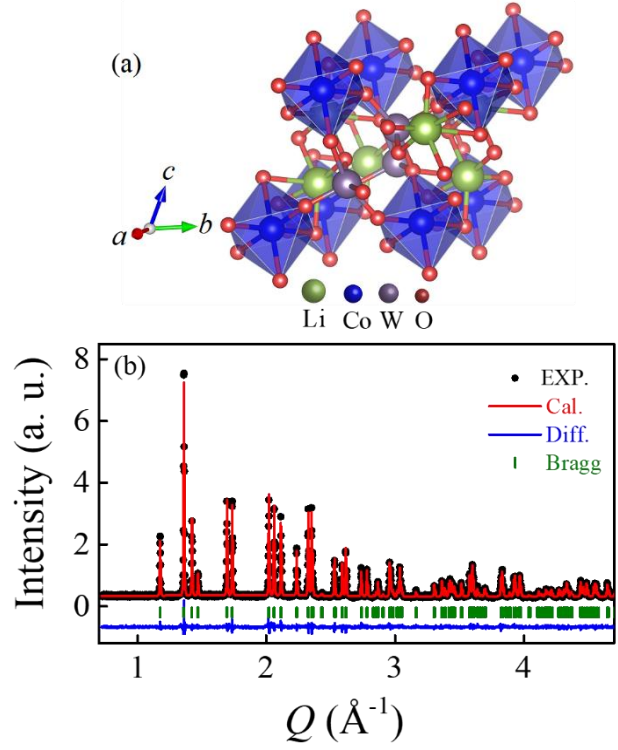

**Fig. 1.** (a) The crystal structure of Li$_2$Co(WO$_4$)$_2$. (b) The measured (black) XRD patterns and the refinement (red) of Li$_2$Co(WO$_4$)$_2$ measured at 300 K. The olive bars denote the Bragg positions and the blue curve shows the difference.

**Table 1.** Refined structural information of Li$_2$Co(WO$_4$)$_2$ from powder X-ray diffraction

| Atom (site) | $x$ | $y$ | $z$ | Occ. | $U_{iso}$ |
|---|---|---|---|---|---|
| Li1 (2i) | 0.0462(2) | 0.0823(3) | 0.2827(2) | 1.000 | 0.0610(8) |
| Co1 (1d) | 0.500 | 0.000 | 0.000 | 1.000 | 0.0237(5) |
| W1 (2i) | 0.2707(1) | 0.5323(5) | 0.6647(3) | 1.000 | 0.0212(3) |
| O1 (2i) | 0.8438(0) | 0.3016(7) | 0.6587(9) | 1.000 | 0.0507(8) |
| O2 (2i) | 0.2271(2) | 0.7393(7) | 0.8015(4) | 1.000 | 0.0605(1) |
| O3 (2i) | 0.6694(9) | 0.6925(0) | 0.5560(1) | 1.000 | 0.0482(2) |
| O4 (2i) | 0.2806(7) | 0.2598(5) | 0.9271(0) | 1.000 | 0.0387(4) |

The temperature dependent magnetic susceptibility $\chi(T)$ of Li$_2$Co(WO$_4$)$_2$ measured under $H = 1$ T is shown in Fig. 2(a). The Curie-Weiss temperature $\theta_{CW} \sim$ -

37.16 K was acquired by fitting the $1/\chi(T)$ curve above 150 K using the Curie-Weiss law $\chi = C/(T - \theta_{CW})$. The negative $\theta_{CW}$ denotes that AFM interactions dominate between $Co^{2+}$ spins. The effective magnetic moment is calculated to be $\mu_{eff} = (8C)^{1/2} = 5.48$ $\mu_B$. This value is larger than expected one 3.87 $\mu_B$ ($S = 3/2$) for high-spin $Co^{2+}$, suggests the existence of the orbital contribution. This value is similar to previous reports for high-spin $S = 3/2$ $Co^{2+}$, e.g., $Co_3Al_2Si_3O_{12}$,[42] $Co_4Nb_2O_9$,[43] $Na_2BaCo(PO_4)_2$,[21] and $BaCo_2(AsO_4)_2$.[22] This high-spin state can persist to low temperatures (e.g. down to 15 K), according to the Curie-Weiss fitting as shown in insert of Fig. 2(a). Interestingly, the $S = 1/2$ low-spin state existing in many cobalt oxides (e.g. $Ba_3CoNb_2O_9$[15] and $Na_2BaCo(PO_4)_2$[21]) at low temperatures is not observed here, which will be further confirmed by the neutron study and DFT calculation.

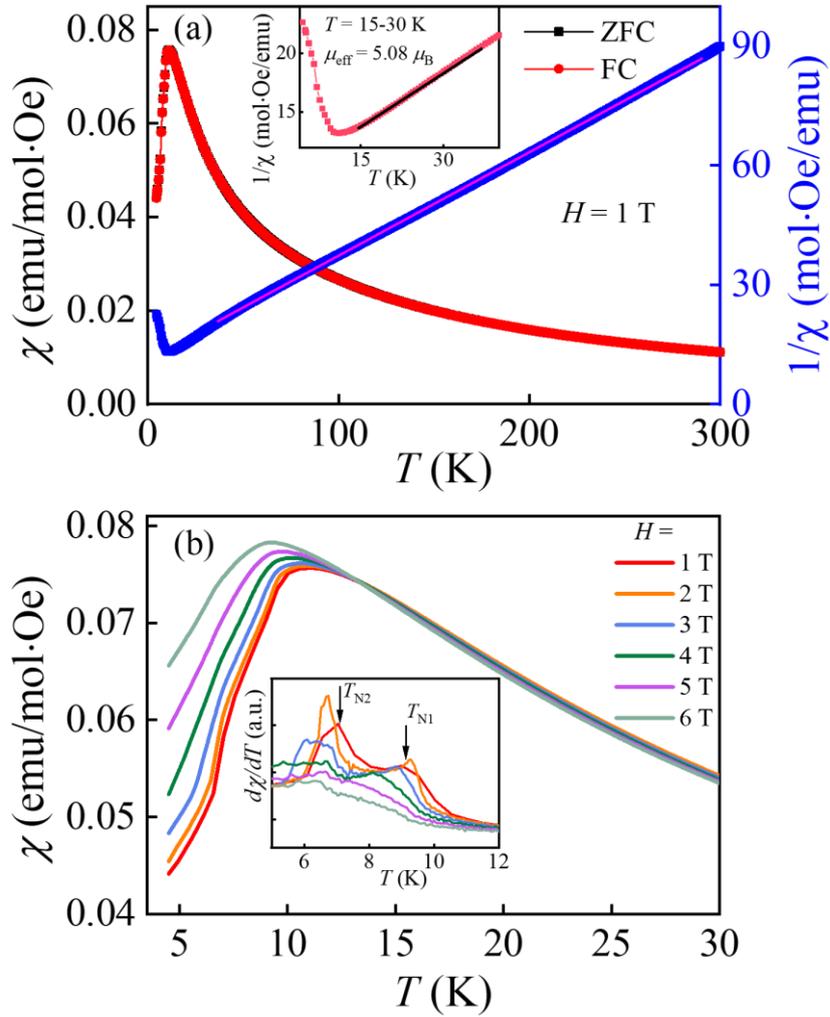

**Fig. 2.** (a) The temperature dependence of magnetic susceptibility $\chi(T)$ (left $y$-axis)

and its inverse (right y-axis) of Li$_2$Co(WO$_4$)$_2$ under $H = 1$ T. Insert: the low temperature range fitting. (b) Magnetic susceptibility $\chi(T)$ under various magnetic fields. The inset displays $d\chi/dT$ around the phase transition temperature range.

Figure 2(b) shows the $\chi(T)$ of Li$_2$Co(WO$_4$)$_2$ measured under various magnetic fields. A broad peak around $T \sim 11$ K is observed originating from the short-range magnetic ordering in low-dimensional magnets. No remarkable anomalies were observed below 10 K. Using the derivative ($d\chi/dT$), two successive peaks at $T_{N1} \sim 9$ K and $T_{N2} \sim 7$ K were clearly seen, as shown in the inset of Fig. 2(b). The broad peak shifts to lower temperature with increasing $H$, consistent with previous studies.[26]

The $H$-dependent isothermal magnetization $M(H)$ at different temperatures, as shown in Fig. 3. The $M(H)$ shows nonlinear behavior under external fields below $T_{N2}$, suggesting possible field-induced transitions such as spin-flop effects.[26,44] Above $T_{N1}$, $M(H)$ show linear behavior and does not saturate up to 9 T.

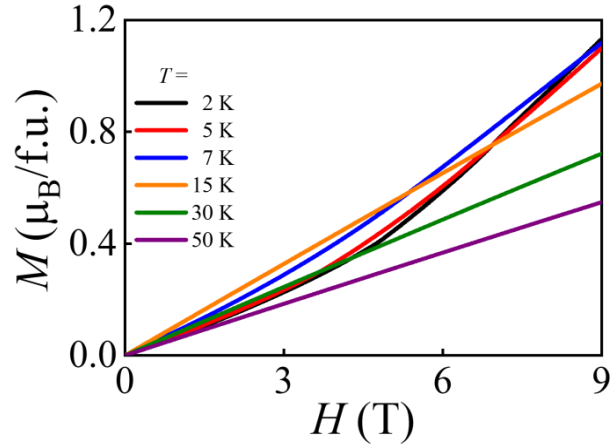

**Fig. 3.** The isothermal magnetization $M(H)$ versus magnetic field at various temperatures. Below T$_{N2}$, the $M(H)$ curves have slope changes at $\sim 5$ T, due to possible spin-flop effects.

Specific heat ($C_p$) of Li$_2$Co(WO$_4$)$_2$ was measured a under different external fields. As shown in Fig. 4(a), two distinct $\lambda$-shaped peaks at $T_{N1} \sim 9.4$ K and $T_{N2} \sim 7.4$ K are observed under zero field, which further confirms that the two successive AFM transitions. Fig. 4(b) shows the $C_p$ of Li$_2$Co(WO$_4$)$_2$ under the selected fields. Both peaks in $C_p$ shift to lower temperature and becomes broader, symbolize the AFM nature. This feature was also observed in other Co-based low-dimensional magnets $ACo_2V_2O_8$ ($A$ = Sr and Ba) and $Ba_3CoNb_2O_9$.[16,17,45]

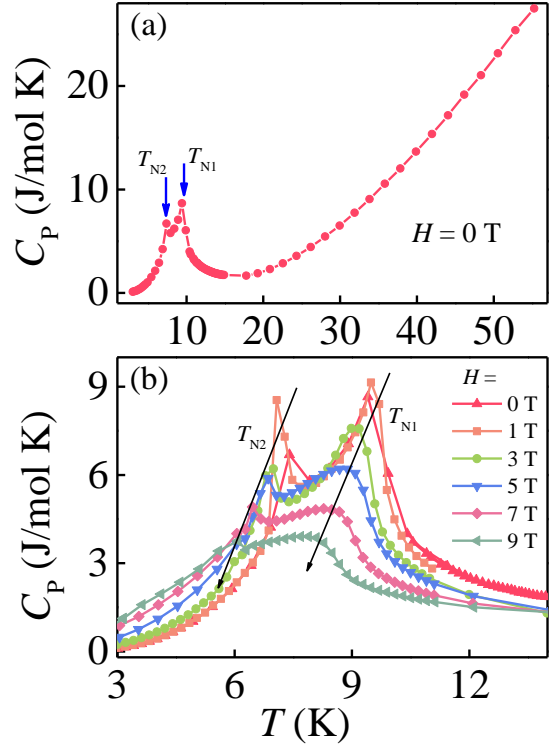

**Fig. 4.** (a) The specific heat ($C_p$) of $Li_2Co(WO_4)_2$ under zero magnetic field. (b) $C_p$ of $Li_2Co(WO_4)_2$ under different applied magnetic fields.

To investigate the magnetic ordering of $Co^{2+}$ ions in $Li_2Co(WO_4)_2$, NPD patterns have been collected both above ($T \sim 30$ K) and below (1.7 K) magnetic transition temperatures. The experimental data (black dots) and the Rietveld refinements profiles (red curves) are shown in Fig. 5(a) and 5(b). The Bragg positions are marked using the vertical bars while the difference between the experimental and refined data is plotted in the insert panel. The nuclear refinement using the 30 K data shows that the lattice parameters are $a = 4.90451(9)$ Å, $b = 5.65117(14)$ Å, $c = 5.86421(12)$ Å, $\alpha = 69.5307(10)°$, $\beta = 91.3690(12)°$ and $\gamma = 116.1986(16)°$, which are in good agreement with our XRD results and previous data.[41]

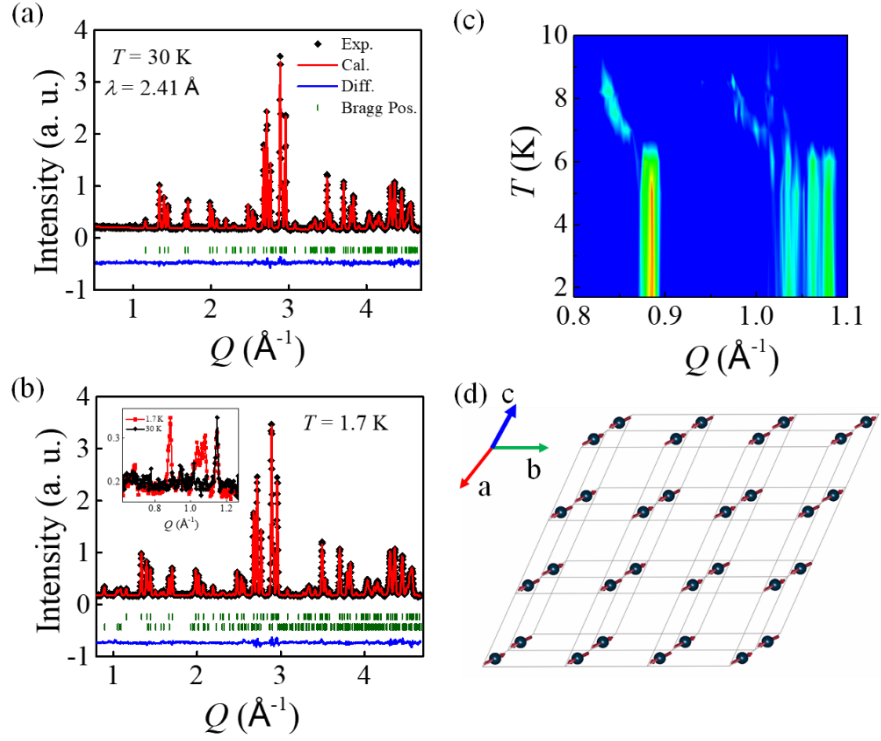

**Fig. 5.** The experimental NPD data collected at $T$ = 30 K (a) and 1.7 K (b) and the refinement results, respectively. The inset displays the enlarged area of the low-$Q$ region where additional magnetic peaks appear. (c) The temperature evolution of the magnetic scattering measured at low-$Q$. (d) The refined magnetic structure of $Li_2Co(WO_4)_2$ at 1.7 K.

At 1.7 K, magnetic peaks emerge as shown in the inset of Fig. 5(b). All peaks were indexed using the propagation vector $\mathbf{k}$ = (½, ¼, ¼), by employing the "k-search" program included in the FullProf package. Considering the $P$-1 symmetry and that there is a single Co atom inside the chemical unit cell, the spin configuration corresponding to the $\mathbf{k}$ = (½, ¼, ¼) is straightforward to evaluate. The doubling of the unit cell along the $a$-direction ($k_a$ = ½) implies an alternation of the spin direction along that axis. However, the moment distribution in the case of $k$ = ¼, as it occurs for the other two crystallographic directions, can be described as $\mathbf{m}_l = \mathbf{m}_o \cos\left(\frac{2\pi}{4} R_l + \varphi\right)$, where $\mathbf{R}_l$ denotes the translation vector of the $l$th unit cell with respect to the zero*th* cell, $\mathbf{m}_o$ is the magnetic in the zero*th* cell, and φ is a phase factor. For a choice of phase φ = 0, the magnetic structure consists of a an amplitude modulated (spin-density-like) sequence ($m_o$, 0, − $m_o$, 0), while for φ = (2n + 1)π/4 the sequence becomes ($m_o$/√2, $m_o$/√2, − $m_o$/√2, − $m_o$/√2). To select between the (+, 0, −, 0) or (+, +, −, −) magnetic structure models one needs to compare the moment values (scale by √2 factor) and determine which one is more physical possible. Our refinements indicate that the magnetic moment is align almost parallel to the $a$-direction. For the case of a uniform moment distribution the refined moment components are $m_a$ = 2.5(1), $m_b$ = - 0.4(1), $m_c$ = - 0.1(1), resulting in a total magnitude $m$ = 2.8(1) $\mu_B$. Alternatively, for an amplitude modulated model the magnetic moment becomes 3.9 $\mu_B$, which appears to be inconsistent with a spin value $S$ = 3/2. The $k$ = (½, ¼, ¼) magnetic structure model yields an excellent fitting result, as shown in the main panel of Fig. 5(b). The generated uniform moment antiferromagnetic structure is plotted in Fig. 5(d). In contrast to $Li_2NiW_2O_8$ and $Li_2Cu(WO_4)_2$,[27,29] the magnetic structure of $Li_2Co(WO_4)_2$ is very unique. Along the $b$ and $c$ directions, we find alternating antiferromagnetic-ferromagnetic interactions [Fig. 5(d)]. We note that our magnetic structure model appears to indicate that contrary to what has been suggested in Ref 19, the $J_1$ exchange integration, corresponding to the shortest Co-Co distance along the $a$-axis, is the strongest coupling; however, there are competing nearest and next nearest couplings along the $b$ and $c$ crystallographic directions that stabilize the alternating AFM-FM spin arrangement.

The temperature evolution of the magnetic peaks was obtained by collecting diffraction patterns using small temperature steps between 2 K and 9 K. The corresponding contour map was displayed in Fig. 5(c). A clear feature is that below $T_{N2}$ ~ 7 K, the magnet Bragg peak positions do not shift as a function of temperature, and the associated propagation vector $k$ is commensurate. In comparison, between $T_{N1}$ and $T_{N2}$, the position of the (½, ¼, ¼) magnetic Bragg peaks located near 0.9 Å$^{-1}$ displays a strong temperature dependence while the intensity becomes weaker and disappears above $T_{N1}$. This is similar to that in $Li_2NiW_2O_8$, three magnetic Bragg peaks appear between $T_{N1}$ and $T_{N2}$, corresponding to an incommensurate SDW-type ordering.[27] It is thus plausible that a similar incommensurate-commensurate magnetic transition also occurs in our Co-based system. The shift in peak position corresponds to a gradual change from the commensurate $k_C$ = (½, ¼, ¼) to an incommensurate $k_{IC}$ = (½, ζ, ξ) wave-vector indicative of an SDW-type ordering.

To further verify the experimental observed magnetic configurations, here a DFT calculation is performed. Three possible magnetic orders are considered, including the ferromagnetic (FM), G-type antiferromagnetic (G-AFM), as well as the one found by neutron study (N-AFM) with a propagation vector $k$ = (½, ¼, ¼). Using the standard GGA+$U$ ($U_{eff}$ = 4 eV) calculation, the energy of N-AFM is the lowest among these three configurations, in agreement with the result of NPD. In addition, the local magnetic moments of $Co^{2+}$ were found to be 2.74 $\mu_B$, also very close to the aforementioned one obtained by NPD. Such a large magnetic moment is about three times of the expected value of $S$ = 1/2 state (e.g. 0.97(1) $\mu_B$/$Co^{2+}$ in $Ba_3CoNb_2O_9$),[15] further supporting the $3d^7$ high-spin configuration. The energy difference between N-AFM and G-AFM is only 0.4 meV/Co. Such a tiny difference implies the low $T_N$, in consistent with the experimental one. Also, the optimized lattice constants ($a$ = 4.924(2) Å, $b$ = 5.620(2) Å, $c$ = 5.879(1) Å) are very close to the experimental one ($a$ = 4.9247(03) Å, $b$ = 5.6707(38) Å, $c$ = 5.8858(63) Å).

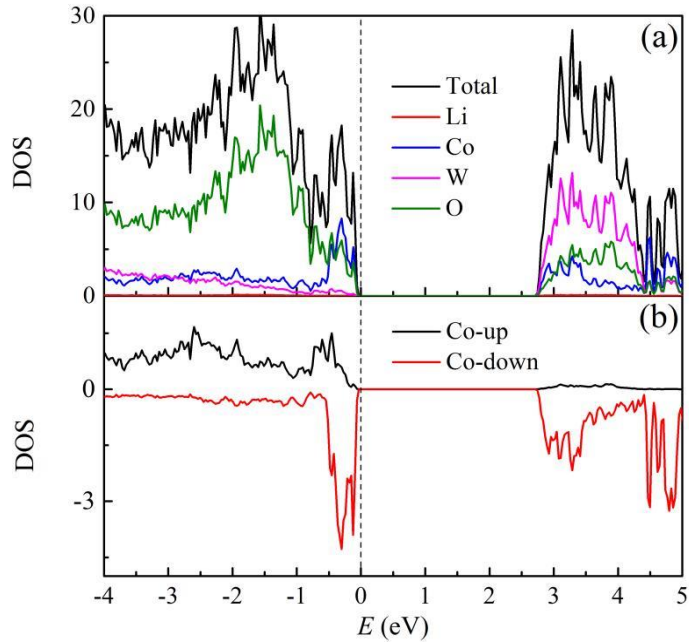

**Fig. 6.** The density of states of $Li_2Co(WO_4)_2$ in the magnetic ground state. (a) The total and atomic projected density of states. (b) Spin-polarized density of states of one

spin-up Co ion.

Then, the electronic structure of $Li_2Co(WO_4)_2$ at the magnetic ground state was calculated, as shown in Fig. 6. $Li_2Co(WO_4)_2$ is an antiferromagnetic insulator with a DFT band gap of $E_g \sim 2.6$ eV, which substantially agrees with the band gap about 3.1 eV measured by light absorption.[46] The atomic projected densities of states are also shown in Fig. 6(a). The topmost valence bands are mostly contributed by the Co and O, while the lowest conducting bands are mostly from W. The spin-polarized density of states shown in Fig. 6(b) suggests the magnetic moments are from $Co^{2+}$ ions, as expected.

## 4. Conclusion

In summary, the magnetic susceptibility, specific heat and neutron diffraction measurements reveal that $Li_2Co(WO_4)_2$ undergoes two successive magnetic transitions. It firstly enters the SDW state at $T_{N1} \sim 9.4$ K, and orders below $T_{N2} \sim 7.4$ K with a commensurate AFM structure characterized by the propagation vector $\bm{k} = $ (½, ¼, ¼). The refined $Co^{2+}$ magnetic moment 2.8(1) $\mu_B$ at 1.7 K. DFT calculation is consistent with the refined magnetic structure and $Co^{2+}$ magnetic moment.

*Note added.* Meanwhile, we note that Karna et al. conducted a similar neutron diffraction study on $Li_2Co(WO_4)_2$, which gave rise to the same conclusion.[47]

**Acknowledgment**

The research at Oak Ridge National Laboratory's High Flux Isotope Reactor was sponsored by the Scientific User Facilities Division, Office of Basic Energy Sciences, US Department of Energy. Most calculations were supported by National Supercomputer Center in Guangzhou (Tianhe II).

## References

[1] Zhou H D, Choi E S, Li G, Balicas L, Wiebe C R, Qiu Y, Copley J R, and Gardner J S 2011 *Phys. Rev. Lett.* **106** 147204

[2] Hase M, Terasaki I I, and Uchinokura K 1993 *Phys. Rev. Lett.* **70** 3651

[3] Imry Y, Pincus P, and Scalapino D 1975 *Phys. Rev. B* **12** 1978

[4] Park S, Choi Y J, Zhang C L, and Cheong S W 2007 *Phys. Rev. Lett.* **98** 057601

[5] Vasiliev A, Volkova O, Zvereva E, and Markina M 2018 *npj Quantum Mater.* **3** 18

[6] Mermin N D and Wagner H 1966 *Phys. Rev. Lett.* **17** 1133

[7] Karmakar K, Skoulatos M, Prando G, Roessli B, Stuhr U, Hammerath F, Ruegg C, and Singh S 2017 *Phys. Rev. Lett.* **118** 107201

[8] Balz C, Lake B, Luetkens H, Baines C, Guidi T, Abdel-Hafiez M, Wolter A U B, Büchner B, Morozov I V, Deeva E B, Volkova O S, and Vasiliev A N 2014 *Phys. Rev. B* **90** 060409(R)

[9] Bera A K, Lake B, Islam A T M N, Klemke B, Faulhaber E, and Law J M 2013 *Phys. Rev. B* **87** 224423

[10] Ma J, Dela Cruz C D, Hong T, Tian W, Aczel A A, Chi S, Yan J Q, Dun Z L, Zhou H D, and Matsuda M 2013 *Phys. Rev. B* **88** 144405

[11] Knafo W, Meingast C, Grube K, Drobnik S, Popovich P, Schweiss P, Adelmann P, Wolf T, and H V L 2007 *Phys. Rev. Lett.* **99** 137206

[12] Moessner R and Ramirez A P 2006 *Phys. Today* **59** 24

[13] Ma Z, Ran K, Wang J, Bao S, Cai Z, Li S, and Wen J 2018 *Chinese Physics B* **27** 106101

[14] Tang Y S, Wang S M, Lin L, Li C, Zheng S H, Li C F, Zhang J H, Yan Z B, Jiang X P, and Liu J M 2019 *Phys. Rev. B* **100** 134112

[15] Lee M, Hwang J, Choi E S, Ma J, Dela Cruz C R, Zhu M, Ke X, Dun Z L, and Zhou H D 2014 *Phys. Rev. B* **89** 104420

[16] He Z, Taniyama T, Kyômen T, and Itoh M 2005 *Phys. Rev. B* **72** 172403

[17] He Z, Taniyama T, and Itoh M 2006 *Phys. Rev. B* **73** 212406

[18] Wang Z, Lorenz T, Gorbunov D I, Cong P T, Kohama Y, Niesen S, Breunig O, Engelmayer J, Herman A, Wu J, Kindo K, Wosnitza J, Zherlitsyn S, and Loidl A 2018 *Phys. Rev. Lett.* **120** 207205

[19] Cui Y, Zou H, Xi N, He Z, Yang Y X, Shu L, Zhang G H, Hu Z, Chen T, Yu R, Wu J, and Yu W 2019 *Phys. Rev. Lett.* **123** 067203

[20] Zhou H D, Xu C, Hallas A M, Silverstein H J, Wiebe C R, Umegaki I, Yan J Q, Murphy T P, Park J H, Qiu Y, Copley J R, Gardner J S, and Takano Y 2012 *Phys. Rev. Lett.* **109** 267206

[21] Zhong R, Guo S, Xu G, Xu Z, and Cava R J 2019 *Proc. Natl. Acad. Sci.* **116** 14505

[22] Zhong R, Gao T, Ong N P, and Cava R J 2020 *Sci. Adv.* **6** eaay6953

[23] Dai J, Zhou P, Wang P-S, Pang F, Munsie T J, Luke G M, Zhang J-S, and Yu W-Q 2015 *Chinese Physics B* **24** 127508

[24] Zhong R, Guo S, Nguyen L T, and Cava R J 2020 *Phys. Rev. B* **102** 224430

[25] Duan L, Wang X-C, Zhang J, Zhao J-F, Cao L-P, Li W-M, Yu R-Z, Deng Z, and Jin C-Q 2020 *Chinese Physics B* **29** 036102

[26] Muthuselvam I P, Sankar R, Ushakov A V, Rao G N, Streltsov S V, and Chou F C 2014 *Phys. Rev. B* **90** 174430

[27] Ranjith K M, Nath R, Majumder M, Kasinathan D, Skoulatos M, Keller L, Skourski Y, Baenitz M, and Tsirlin A A 2016 *Phys. Rev. B* **94** 014415

[28] Muthuselvam I P, Sankar R, Singh V N, Rao G N, Lee W L, Guo G Y, and Chou F C 2015 *Inorg. Chem.* **54** 4303

[29] Ranjith K M, Nath R, Skoulatos M, Keller L, Kasinathan D, Skourski Y, and Tsirlin A A 2015 *Phys. Rev. B* **92** 094426

[30] Dergachev K G, Kobets M I, and Khatsko E N 2005 *Low Temp. Phys.* **31** 402

[31] Nyam-Ochir L, Ehrenberg H, Buchsteiner A, Senyshyn A, Fuess H, and Sangaa D 2008 *J. Magn. Magn. Mater.* **320** 3251

[32] Liu M, Lin L, Zhang Y, Li S, Huang Q, Garlea V O, Zou T, Xie Y, Wang Y, Lu C, Yang L, Yan Z, Wang X, Dong S, and Liu J-M 2017 *Phys. Rev. B* **95** 195134

[33] Zhao D, Shi J-C, Nie C-K, and Zhang R-J 2017 *Optik* **138** 476

[34] Zapart W and Zapart M B 2016 *Ferroelectrics* **497** 126

[35] Liu M, Zhang Y, Zou T, Garlea V O, Charlton T, Wang Y, Liu F, Xie Y, Li X, Yang L, Li B, Wang X, Dong S, and Liu J M 2020 *Inorg. Chem.* **59** 8127

[36] Kresse G and Furthmüller J 1996 *Comput. Mater. Sci.* **6** 15

[37] Kresse G and Furthmüller J 1996 *Phys. Rev. B* **54** 11169

[38] Perdew J P, Ruzsinszky A, Csonka G I, Vydrov O A, Scuseria G E, Constantin L A, Zhou X, and Burke K 2008 *Phys. Rev. Lett.* **100** 136406

[39] Gao B, Lin L, Chen C, Wei L, Wang J, Xu B, Li C, Bian J, Dong S, Du J, and Xu Q 2018 *Phys. Rev. Mater.* **2** 084401

[40] Dudarev S L, Botton G A, Savrasov S Y, Humphreys C J, and Sutton A P 1998 *Phys. Rev. B* **57** 1505

[41] Álvarez-Vega M, Rodríguez-Carvajal J, Reyes-Cárdenas J G, Fuentes A F, and Amador U 2001 *Chem. Mater.* **13** 3871

[42] Cui Q, Huang Q, Alonso J A, Sheptyakov D, De la Cruz C R, Fernández-Díaz M T, Wang N N, Cai Y Q, Li D, Dong X L, Zhou H D, and Cheng J G 2020 *Phys. Rev. B* **101** 144424

[43] Khanh N D, Abe N, Sagayama H, Nakao A, Hanashima T, Kiyanagi R, Tokunaga Y, and Arima T 2016 *Phys. Rev. B* **93** 075117

[44] Wang G H, Xu C Y, Cao H B, Hong T, Huang Q, Ren Q Y, Xu J Q, Zhou H D, Luo W, Qian D, and Ma J 2019 *Phys. Rev. B* **100** 035131

[45] Yokota K, Kurita N, and Tanaka H 2014 *Phys. Rev. B* **90** 014403

[46] Karoui K, Rhaiem A B, and Jemni F 2021 *Ionics* **27** 1511

[47] Karna S K, Wang C W, Sankar R, Temple D, and Avdeev M 2021 *Phys. Rev. B* **104** 134435